\documentclass[a4paper]{jpconf}
\usepackage{graphicx}
\usepackage[square,sort&compress]{natbib}

\def\kms{\ifmmode{\rm km\thinspace s^{-1}}\else km\thinspace s$^{-1}$\fi}
\def\lamsco{$\lambda$~Sco}
\def\pisco{$\pi$~Sco}

\def\psicen{$\psi$~Cen}
\def\mueri{$\mu$~Eri}
\def\betaaur{$\beta$~Aur}
\def\deltacap{$\delta$~Cap}
\def\arcas{AR~Cas}
\def\acen{$\alpha$~Cen~A}
\def\altair{Altair}
\def\epscep{$\epsilon$~Cep}
\def\cyg29{29~Cyg}

\def\wire{WIRE}
\def\hipp{HIPPARCOS}

\def\teff{$T_{\rm eff}$}

\def\ctopp{\emph{Top}}
\def\cmidd{\emph{Middle}}
\def\topp{\emph{top}}
\def\midd{\emph{middle}}
\def\cbott{\emph{Bottom}}
\def\bott{\emph{bottom}}

\def\panel{\emph{panel}}
\def\panels{\emph{panels}}

\def\dss{$\delta$~Scuti}
\def\betacep{$\beta$~Cep}
\def\str{Str\"omgren}

\begin{document}
\title{A new level of photometric precision:\\ WIRE observations of eclipsing binary stars}

\author{Hans Bruntt$^1$ and John Southworth$^2$}

\address{$^1$School of Physics A28, University of Sydney, NSW 2006, Australia}
\address{$^2$Department of Physics, University of Warwick, Coventry, CV4 7AL, UK}

\ead{bruntt@physics.usyd.edu.au, jkt@astro.keele.ac.uk}

\begin{abstract}
The \wire\ satellite was launched in March 1999 and was the first space mission to do asteroseismology from space on a large number of stars. \wire\ has produced very high-precision photometry of a few hundred bright stars ($V<6$) with temporal coverage of several weeks, including K~giants, solar-like stars, \dss\ stars, and $\beta$~Cepheids. In the current work we will describe the status of science done on seven detached eclipsing binary systems. Our results emphasize some of the challenges and exciting results expected from coming satellite missions like COROT and Kepler. Unfortunately, on 23 October 2006, communication with \wire\ failed after almost eight years in space. Because of this sad news we will give a brief history of \wire\ at the end of this paper.
\end{abstract}


\section{Observing bright stars with the \wire\ star tracker}

The failure of the main mission of the \wire\ satellite shifted focus to the star tracker, which has a small aperture of $52$\,mm and is equipped with a 512$\times$512 pixel SITe CCD. During observations the main target is positioned near the middle of the CCD and the four brightest stars in the 8$\times$8 degree field of view are also monitored. These four secondary stars are chosen automatically by the on-board software. Each observation comprises a time stamp and an 8$\times$8 pixel window centred on the target. Two images per second are collected for each star, resulting in typically one million CCD windows in three weeks. Due to pointing restrictions two fields are observed during each \wire\ orbit. The duty cycle for one star per orbit is optimally 40\%, but can be as low as 20\%. The orbital period has decreased from 96 to 93 minutes over the cause of the mission. The filter reponse of the star tracker is not well defined but Johnson $V+R$ has been suggested \citep{buzasi00}. For more details about \wire\ see Sect.~\ref{sec:epitaph}.

\begin{figure*}[h]
\centering
\includegraphics[width=14cm]{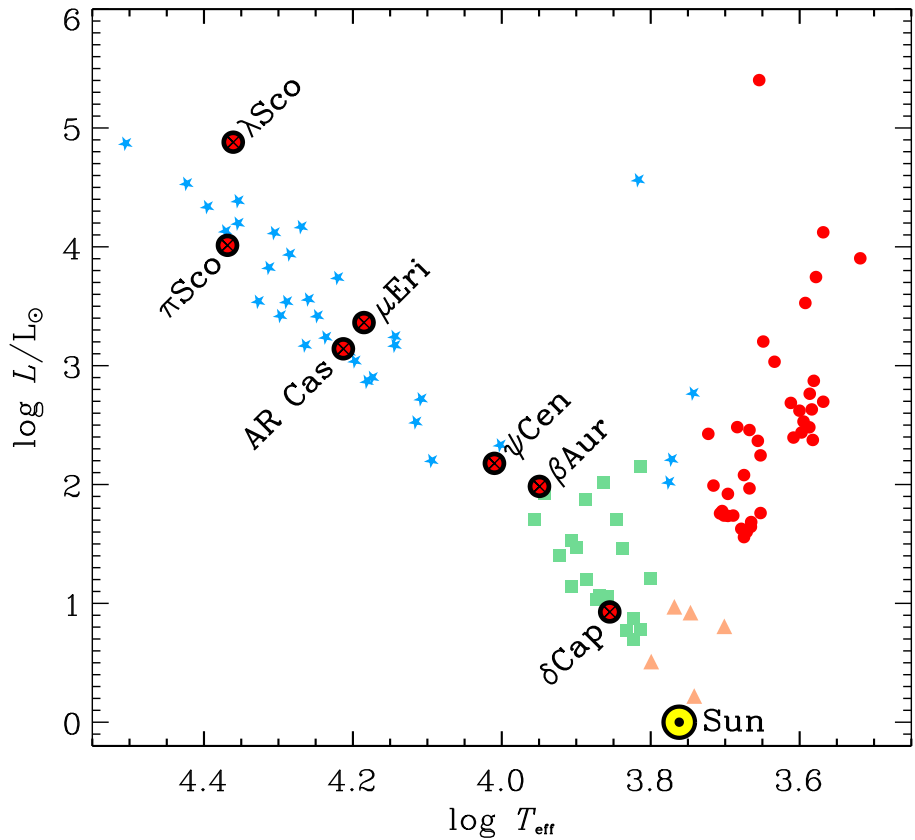}
\caption{\label{fig:hr} Hertzsprung-Russell diagram of around 100 stars
observed with \wire. The locations of the seven dEBs and the Sun are marked.}
\end{figure*}

\section{Eclipsing binary stars observed with \wire}

The location of the 100 \wire\ targets with the best photometry are shown in Fig.~\ref{fig:hr}. The seven detached eclipsing binaries (dEBs) observed by \wire\ are marked, and basic information for these objects is given in Table\,\ref{tab:bin}. For each of these systems the \wire\ light curve is hugely better than existing observations, allowing definitive analyses of even the difficult objects.

\subsection{The seven eclipsing binaries observed with \wire}

Four of the dEBs (\betaaur, \arcas, \deltacap\ and \lamsco) were observed as primary targets by \wire. \mueri\ was a secondary target but is known to be a remarkable object which shows slowly pulsating B-type (spB) star variation and eclipses \cite{Jerzykiewicz+05mn}. The sixth system, \pisco, is probably an ellipsoidal variable. Finally, \psicen\ was discovered to be a dEB from its \wire\ light curve. The fact that we are still discovering eclipses in stars visible to the naked eye illustrates just how much science remains to be done with the brightest and closest stars to Earth.

We are in the process of modelling the light curves of the dEBs observed by \wire\ in order to measure their properties. When combined with radial velocity analyses we will measure the masses and radii of the component stars with an accuracy better than 1\%, which is generally needed to apply detailed constraints to the properties of theoretical stellar evolutionary models \cite{Andersen91aar}.

To analyse the light curves we are using the {\sc jktebop} modelling code\footnote{{\sc jktebop} is written in {\sc fortran77} and the source code is available at {\tt http://www.astro.keele.ac.uk/$\sim$jkt/}} \cite{Me++04mn}, which is based on the {\sc ebop} code written by Etzel \cite{PopperEtzel81aj}.
The major advance of {\sc jktebop} is the implementation of a Monte Carlo error estimation algorithm \cite{Me++04mn2,Me+04mn3,Me+05mn}, which is a vital source of reliable parameter uncertainties when only one light curve is available for a dEB. For our work on the \wire\ light curves we have added several new features, including non-linear limb darkening laws \cite{south07,Me++07mn} and the direct incorporation of spectroscopic light ratios and times of minimum light \cite{south07}.




\begin{table}[b]
\centering
\caption{\label{tab}The seven eclipsing binary stars observed with WIRE.\label{tab:bin}}
\begin{tabular}{l|ll|r|l}
\br
Star & \multicolumn{2}{c|}{Spectral types} & Period [d] & Notes \\
\mr
\deltacap   & Am      & F    &  1.02   & Active component, see \cite{lloyd94} \\
\pisco      & B1V     & B2   &  1.57   & \betacep\ components \\
\betaaur    & A1 m    & A1 m &  3.96   & Published~\cite{south07} \\
\lamsco     & B1.5 IV & B2 V &  5.95   & Triple system \cite{uytter04a} \\
\arcas      & B4 V    & A6 V &  6.06   & spB component \\
\mueri      & B5 IV   & ?    &  7.38   & spB component, see \cite{Jerzykiewicz+05mn} \\
\psicen     & B9 V    & A2 V & 38.81   & Published~\cite{bruntt06} \\
\br
\end{tabular}
\end{table}

\subsection{Further observations of the newly discovered eclipsing binary \psicen}



\begin{figure*}[h]
\centering
\includegraphics[width=15cm]{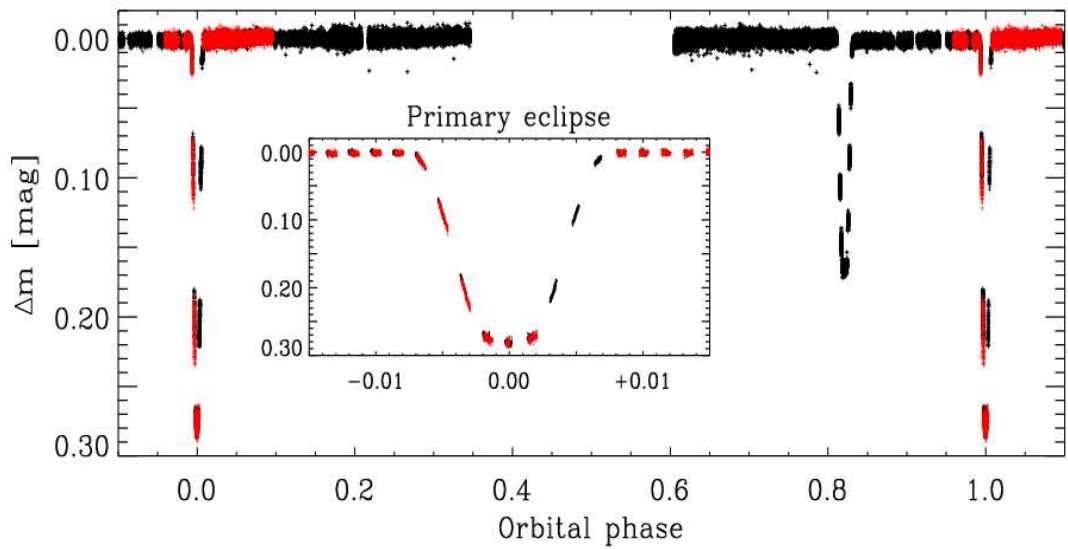}
\caption{\label{fig:psicen} Phased light curve of \psicen\ from \wire.
We show the original time series from 2004 along with the
new data from 2006 in red colour.}
\end{figure*}

In \cite{bruntt06} we presented the discovery and photometric analysis of a new bright ($V = 4.1$) well-detached EB, \psicen, previously known to have a variable radial velocity \cite{BuscombeMorris61mn}. The phased \wire\ light curve in Fig.\,\ref{fig:psicen} shows two deep flat-bottomed eclipses.
\wire\ observed a short additional light curve of a primary eclipse of \psicen\ in 2006 July, allowing us to refine the orbital period to $P=38.811929 \pm 0.000012$\,d.

To further constrain the parameters of the \psicen\ system, we have obtained radial velocity data from FEROS at the ESO~2.2\,m telescope on La~Silla. We have spectra from 33 epochs covering the full orbit and the data are currently being processed. Also, we have obtained Johnson $B,V$ photometry in order to measure the colour difference in the eclipses, which will in turn allow us to constrain the ratio of the \teff s of the two stars.



\subsection{\wire\ observations of \betaaur, the first known double-lined eclipsing binary}

\betaaur\ was the first known double-lined spectroscopic binary, discovered in the course of the {\it Henry Draper Catalogue} project by \cite{Maury90}. It was subsequently found to be eclipsing \cite{Stebbins11apj}, making it also the first known double-lined eclipsing binary. It is very bright ($V = 1.9$), has an orbital period close to four days ($P = 3.9600$\,d), and shallow eclipses which are less than 10\% deep. These characteristics have made it a remarkably difficult object to study, particularly in the current era of rapidly increasing telescope aperture size. Conversely, the similarity and low rotational rates of the two stars mean it is very easy to study spectroscopically. As a result, the best study of \betaaur\ \cite{NordstromJohansen94aa} was based on light curves from the 1960s \cite{Johansen71aa} and a photographic radial velocity analysis from the 1940s \cite{Smith48apj}. \betaaur\ also has a high-precision parallax from \hipp\ \cite{Perryman+97aa} and an orbital parallax from the Mark\,III Optical Interferometer \cite{Hummel+95aj}.

\begin{figure*}[h] \centering
\includegraphics[width=16cm]{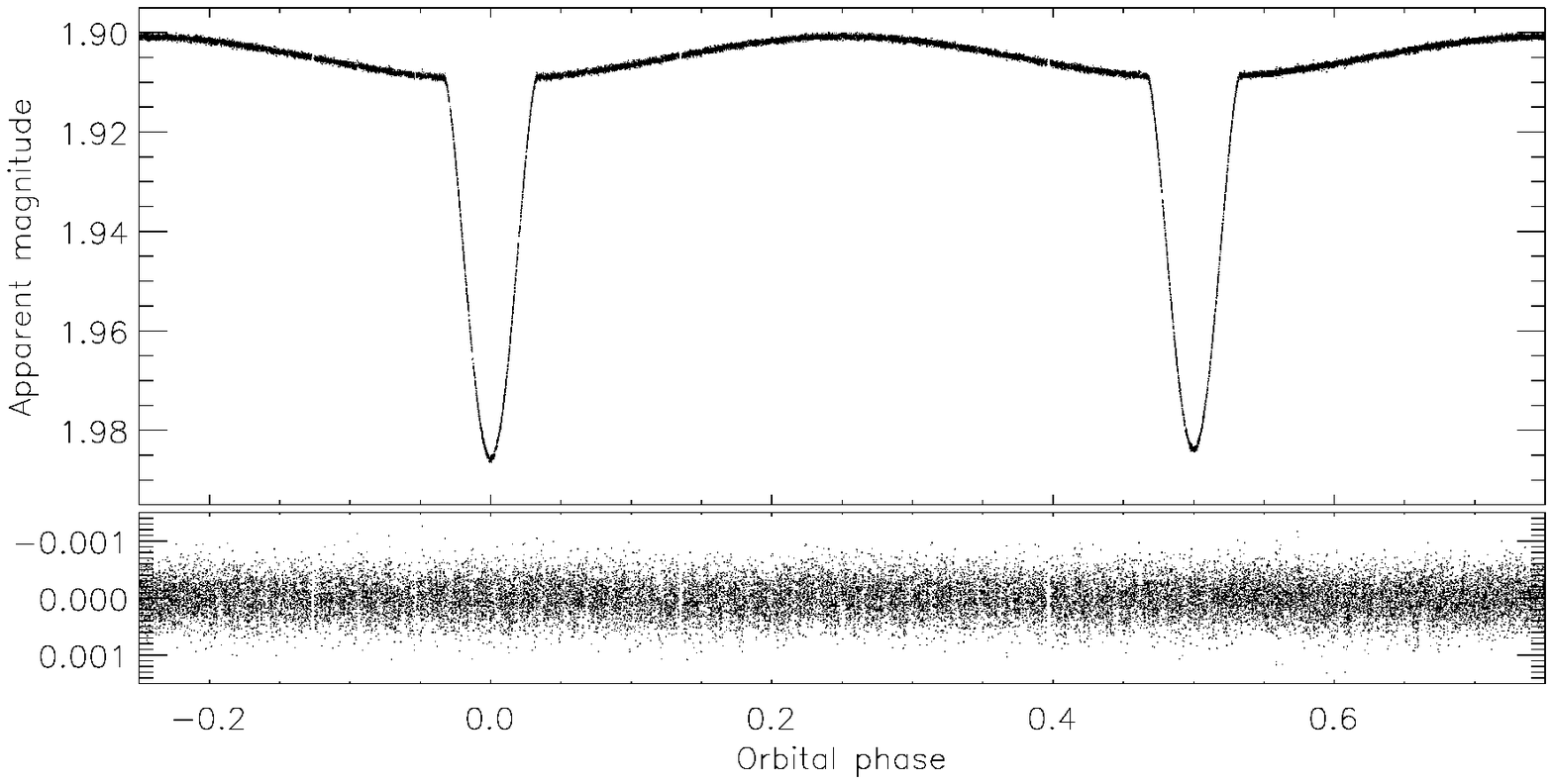}
\caption{\label{fig:baur1} \ctopp\ \panel: phased light curve of \betaaur\ obtained
using \wire. \cbott\ \panel: residuals of the best {\sc jktebop} fit. } \end{figure*}

We obtained a \wire\ light curve (Fig.~\ref{fig:baur1}) covering 21 days in 2006 April and consisting of 30\,015 observations with a point-to-point scatter of only 0.3\,mmag. The characteristics of the light curve required us to modify the {\sc jktebop} code to include spectroscopic light ratios and times of minimum light directly in the solution. We also took the opportunity to measure the coefficients of non-linear limb darkening laws for the first time for an EB system. The best-fitting model is an excellent match to the data (Fig.\,\ref{fig:baur1}) and yields the fractional radii of the two stars to 0.5\%. Using the radial velocity analysis of \cite{Smith48apj} we find masses of $M_{\rm A} = 2.38 \pm 0.03$\,M$_\odot$ and $M_{\rm B} = 2.29 \pm 0.03$\,M$_\odot$, and radii of $R_{\rm A} = 2.76 \pm 0.02$\,R$_\odot$, $R_{\rm B} = 2.57 \pm 0.02$\,R$_\odot$ for the two stars \cite{south07}. The surface brightness method \cite{Me++05aa} gives a distance to \betaaur\ of $25.0 \pm 0.4$\,pc, in very nice agreement with the \hipp\ distance of $25.2 \pm 0.5$\,pc.


\subsection{A new system: \mueri}

\begin{figure*}[h]
\centering
\includegraphics[width=15.0cm]{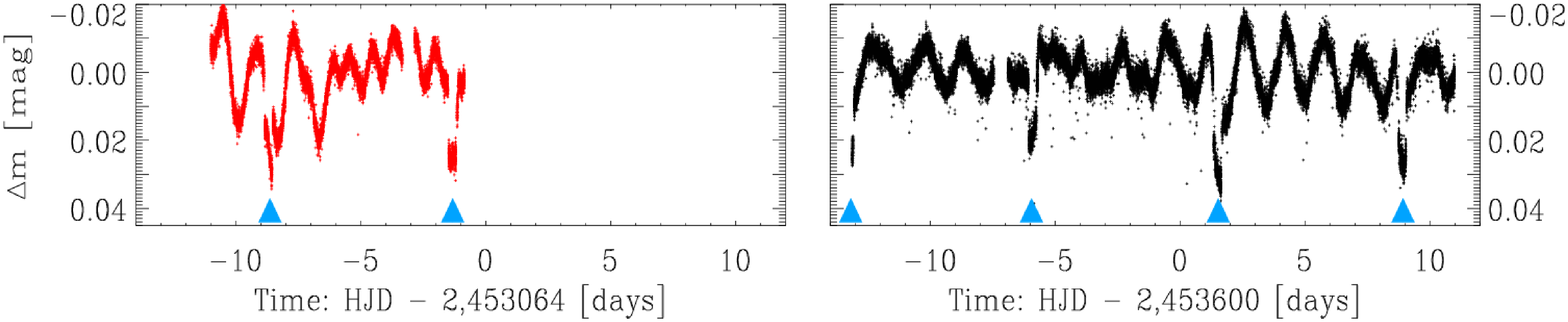}
\includegraphics[width=15.7cm]{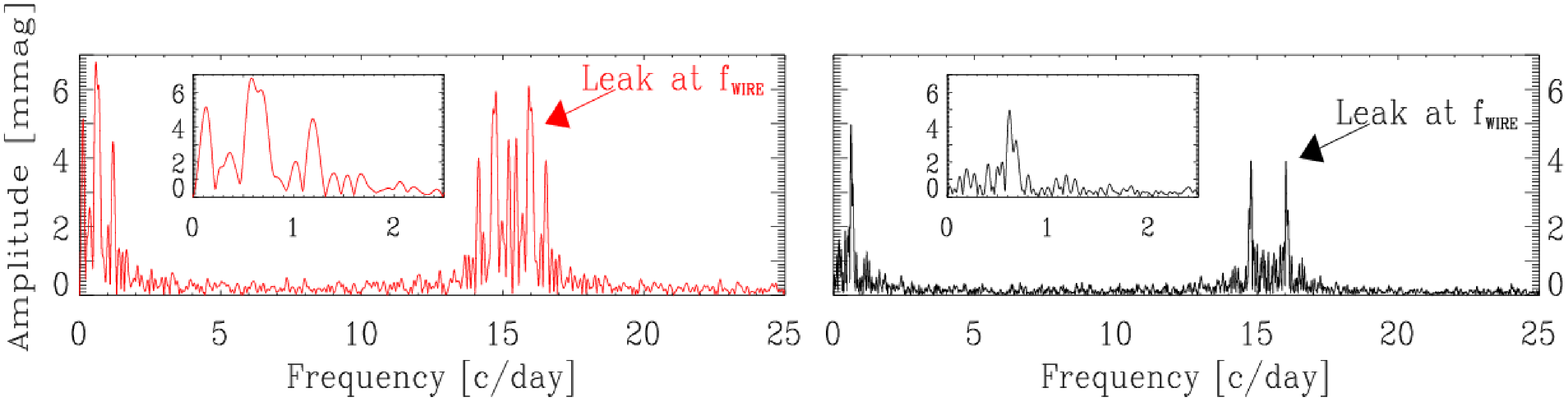}
\includegraphics[width=15.7cm]{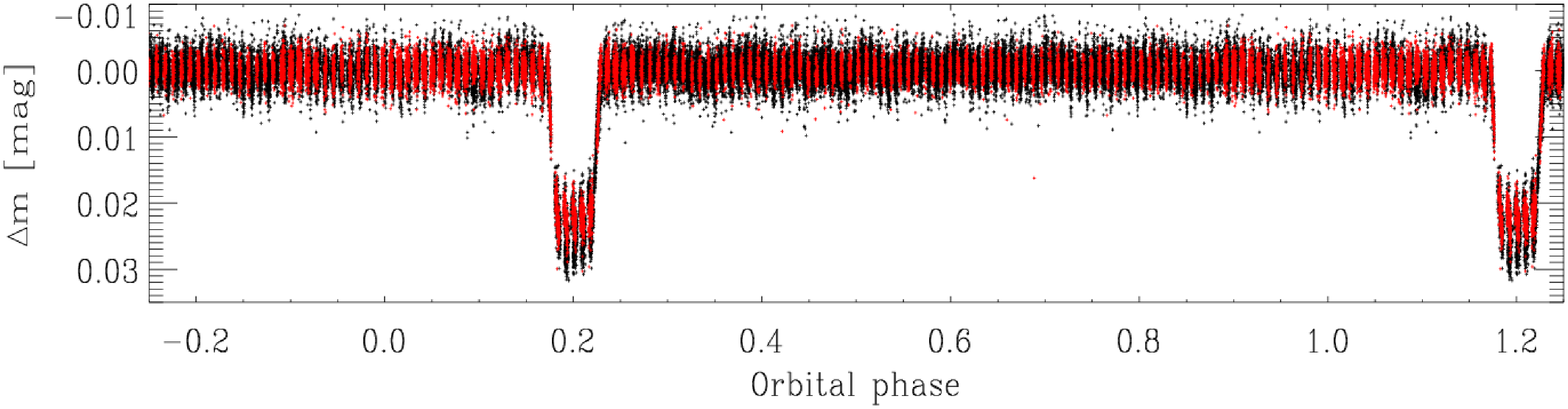} 
\caption{\label{fig:mueri}
\ctopp\ \panels: light curves of \mueri\ from February 2004 (in red) and August 2005.
\cmidd\ \panels: amplitude spectra of the light curves (eclipses removed). Excess power below 1~c/day is seen to
leak to frequencies around the orbital frequency of \wire.
\cbott\ \panels: the phased time series after oscillations have been subtracted.}
\end{figure*}

\mueri\ was found to be an spB star by \cite{Handler+04mn} and subsequently discovered to be eclipsing \cite{Jerzykiewicz+05mn}. It was observed by \wire\ as a secondary target for eight days in 2004 February and 24 days in 2005 August. The light curves are shown in the \topp\ \panels\ in Fig.\,\ref{fig:mueri}: both show variation at timescales of $1.5$--$2.5$ days with beating of several modes. These periods are consistent with spB variations also seen by \cite{Jerzykiewicz+05mn} and in agreement with the spectral type of B5\,IV. In addition, eclipses occur every $\simeq7.4$ days (marked by triangles in Fig.\,\ref{fig:mueri}). In the \midd\ \panels\ we show the amplitude spectra of the light curves (excluding data taken during the eclipses). Note that the variation at low frequencies leak into frequencies near the orbital frequency of \wire\ at $\simeq15.4$\,cycles/day. In the \bott\ \panels\ we shown the two time series with the oscillations removed and phased with the orbital period of 7.381\,d.


We have fitted the detrended light curve with {\sc jktebop} and have confirmed that there is a chance of measuring the radii of the two stars to within a few per cent. However, this brute-force approach will need refining to obtain trustworthy results: at present it can be seen that the eclipse depth is apparently variable in the detrended data (Fig.\,\ref{fig:mueri}). We will need to obtain new radial velocities, and then study these {\em simultaneously} with the oscillations and eclipses. This work is in progress.


\subsection{Preliminary results for \arcas}

\begin{figure*}[h] \centering
\includegraphics[width=15.5cm]{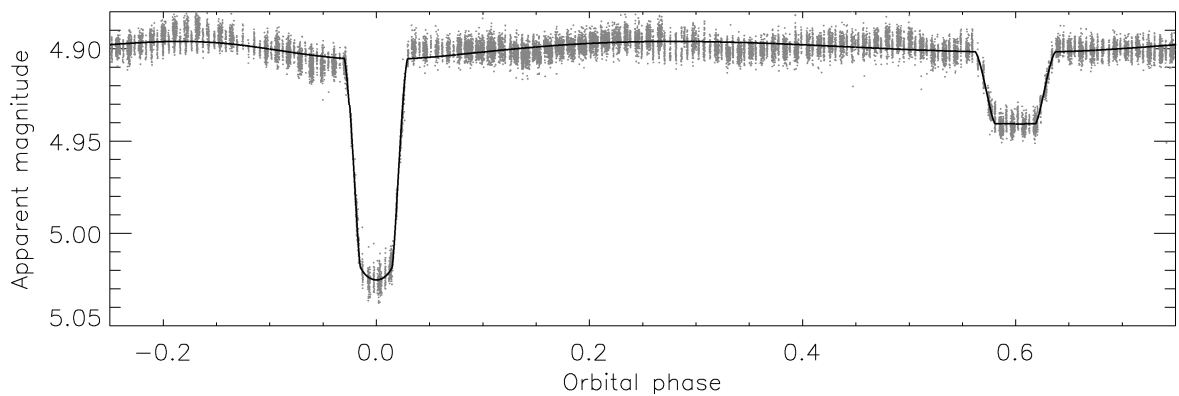}
\caption{\label{fig:arcas1} The phased light curve of \arcas\ (points) is
compared to the best {\sc jktebop} fit (solid line).
The pulsations have not been subtracted from the light curve.} \end{figure*}

AR\,Cassiopeiae is a bright ($V = 4.9$) dEB composed of a B4\,V primary star and a much less massive A6\,V secondary component. \cite{Holmgren+99aa} were the first group to find the secondary star in the spectrum, as this component only contributes a few per cent of the light of the system. \arcas\ has an orbital period of 6.06 days, an eccentric orbit, and exhibits apsidal motion, probably with a long apsidal period \cite{Holmgren+99aa}.
Our \wire\ light curves of \arcas\ show that the system undergoes shallow total eclipses. We have fitted the data with {\sc jktebop} and the phased light curve and fit (solid line) are compared in Fig.\,\ref{fig:arcas1}. The light curve fit is good for the eclipses but is less good outside eclipse as the primary component of \arcas\ is intrinsically variable. A periodogram of the residuals of the best fit shows significant power at several frequencies and suggests that the primary of \arcas\ may be an spB star.

We are in the process of studying \arcas\ both photometrically and spectroscopically. At this point we have obtained over 100 grating spectra using the Isaac Newton Telescope and IDS spectrograph and 27 high-resolution \'echelle spectra with the recently-installed FIES instrument at the Nordic Optical Telescope, and will be obtaining further data in 2007 October. The light curve will need a similar approach to \mueri\ -- simultaneous inclusion of both pulsational and eclipse effects -- and an implementation of this into {\sc jktebop} is in progress.

%

%
%


\subsection{Preliminary results for \deltacap}

\begin{figure*}[h] \centering
\includegraphics[width=15.5cm]{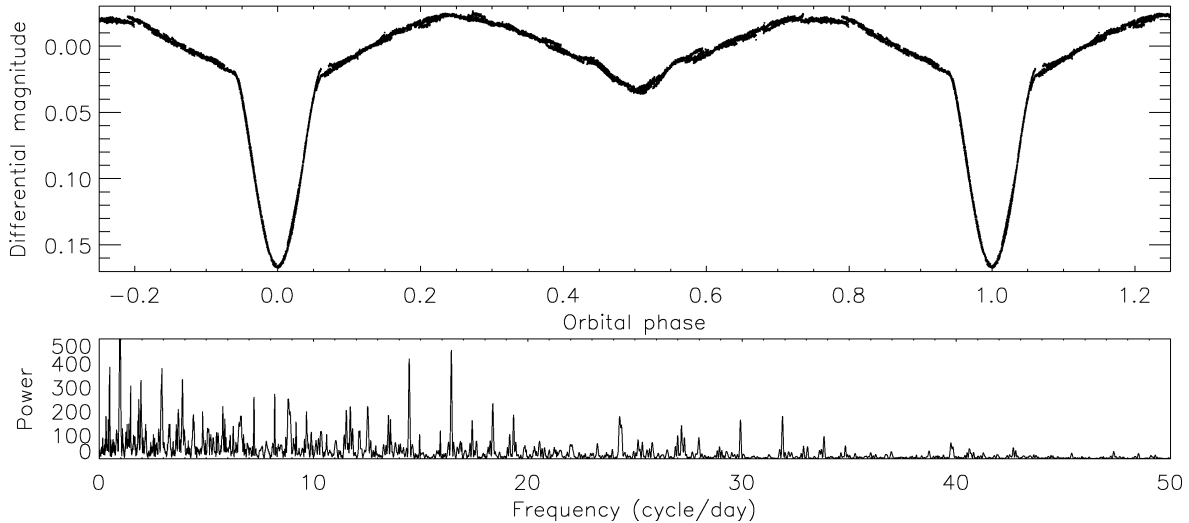}
\caption{\label{fig:dcap} \ctopp\ \panel: the phased \wire\
light curve of \deltacap\ shows clear primary eclipses, very
shallow secondary eclipses, and significant intrinsic variation
arising from the much brighter primary star. \cbott\ \panel:
a periodogram of the residuals of the fit displays a rich power
spectrum characteristic of an active star.} \end{figure*}

The last dataset to be obtained from the \wire\ satellite was a twelve-day light curve of the active star \deltacap\ ($V = 2.9$). The phased light curve shown in Fig.~\ref{fig:dcap} shows shallow eclipses with a ``nasty'' orbital period of 1.02 days, making it nearly impossible to observe from the ground. The primary component is a metallic-lined A-star and the much fainter secondary is probably an F-star. The system is a known X-ray source and was detected by ROSAT \cite{Wonnacott+92mn}. Again, the analysis of this light curve presents serious challenges: the total light contribution from the secondary star is comparable to the amplitude of the brightness changes of the primary star.

\section{Conclusion}

We have presented on-going work on the observations and modelling of several
eclipsing binary systems observed with the \wire\ satellite.
The week-long temporal coverage of the targets has made it possible to study
dEB systems that are notoriously difficult to observe from the ground,
due to their periods being close to an integer number of days.
The high precision of the photometry from 
\wire\ is a huge improvement compared to even the best photometric observations from the ground.
The two important advantages of space based photometry are:
\begin{itemize}
\item No atmospheric scintillation noise
\item High stability of the \wire\ star tracker over periods of several weeks.
\end{itemize}

These data have made it possible to push the limits on the constraints
we can put on the theoretical models of these stars, covering the HR diagram
from early B to late F type main sequence stars. However, at this level of precision the
oscillations of the component stars, although amplitudes are small, need to be taken into account
as part of the light curve analysis.

In a broader context, the results presented here give an idea of the potential
of secondary science on dEBs that can be done with data from the future satellite photometry missions.
Many new systems will be discovered with COROT (see \citep{michel05}) and Kepler as discussed by \cite{koch07}.
Although the new systems will be much fainter, follow-up spectroscopy
and multi-band photometry can still be done from the ground.




\section{WIRE epitaph: eight years of high-precision photometry from space\label{sec:epitaph}}


\begin{figure}[b] \centering
\includegraphics[angle=70,width=10.4cm]{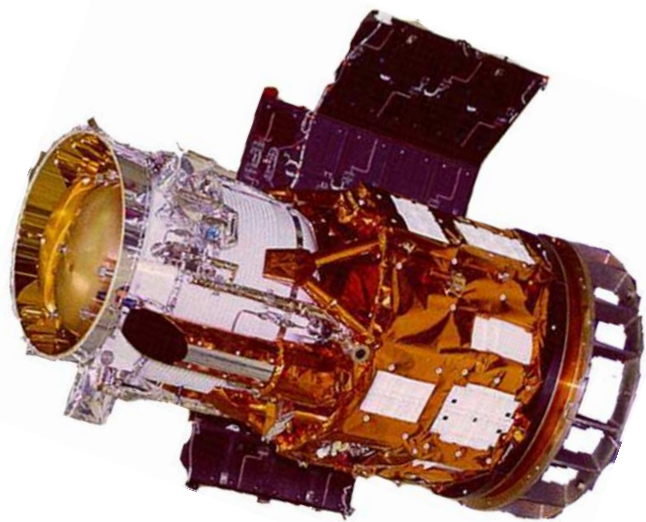}
\caption{\label{fig:wire} The \wire\ satellite with its successful star tracker on the side of it.}
\end{figure}

In 1994, \wire\ was selected as the fifth NASA satellite mission under the Small Explorer (SMEX) program.
\wire\ was successfully launched into a sun-synchronous low-Earth orbit on 4 March 1999 \cite{laher00}.
Equipped with a 12$^{\prime\prime}$ telescope and an infrared camera cooled with hydrogen, it was designed to
study faint star burst galaxies \cite{hacking99}. However, the cover for the main camera opened
three days too soon and the instrument was exposed to sun light. As a result,
the hydrogen coolant was quickly lost, 
and as a consequence the main mission failed. Derek L.\ Buzasi, who was working at NASA
at the time, realized that the fully functional star tracker could potentially be
used to monitor oscillations in bright stars \cite{buzasi00}.
The planned mission lifetime of \wire\ was only four months,
but the secondary asteroseismology mission with the star tracker lasted nearly eight years.

\wire\ was used for two epochs with slightly different observing modus.
In epoch~1 from May 1999 to September 2000 around 25 stars were monitored.
The star tracker kept the prime target on the same position within a few hundredths
of a pixel over periods of several weeks \cite{bruntt07a}.
In addition four other stars were observed, but
due to the slow rotation of the space craft, the four secondary stars drifted across
the CCD pixels. It was not possible to acquire a flat field so variations
of a few per cent were seen in the data with typical time scales of a day.
As a consequence, the secondary targets in epoch~1 were not usable for detailed studies.
Highlights of the research from epoch~1 include the detection of $p$~modes
in the solar-like stars \acen\ \cite{schou01,fletcher06} and Procyon~A \cite{bruntt05},
and the discovery that \altair\ is the brightest \dss\ star in
the sky \cite{buzasi05,suarez05}.
Mainly due to funding problems there was a break for over three years
until the observations started again in late 2003.

Epoch~2 lasted from December 2003 until
23 October 2006 at 18.14 UT, when communication with
the satellite failed. Reprogramming the modus operandi of the satellite
made it possible to keep all five ``tracker stars'' fixed on the same pixel position,
thus avoiding the problem of not having a flat field image.
Consequently, a multitude of secondary targets were observed and today
300 light curves of around 200 stars exist in the \wire\ database \cite{bruntt07a}.
During the second epoch a pipeline was developed to reduce the data \cite{bruntt05}.
In the last two years of operation a few campaigns were organised
where ground-based support was made. For two \dss\ stars, \epscep\ \cite{bruntt07b}
and \cyg29\ (now being analysed), simultaneous \str\ $uvby$ data were collected in
an attempt to identify the oscillation modes by measuring amplitude ratios
and phase differences between different filters. 
Furthermore, large ensembles of pulsating stars of different spectral types were observed.
Around 20 \betacep\ and 15 spB-type pulsators have been monitored (see Fig.~\ref{fig:hr}).
Also, a systematic change with evolutionary status of the amplitude and
frequency in K~giants has recently been reported \cite{stello08}.

The idea of observing the oscillation of stars from space came about in the early 1980s \cite{roxburgh02}.
After the unfortunate loss of the EVRIS mission in 1996, another 10 years passed until
COROT was launched in December 2006 (see Michel, these proceedings).
Before that the WIRE satellite and the Canadian MOST asteroseismology mission
were launched in March 1999 and June 2003, respectively.
The potential promise of great science from photometry from space
was realized somewhat earlier, based on the \hipp\ mission and
the guide stars used by the Hubble Space Telescope \cite{zwintz00}.

The future looks bright for asteroseismology from space with more missions
being planned. NASA's Kepler mission is set for launch in February 2009 \cite{jcd07},
and will observe 512 asteroseismic targets (at high cadence) for up to four years.
A recently proposed ESA mission is Plato (see Catala, these proceedings).
On a slightly smaller scale we mention the SMEI instrument now flying on the
Coriolis mission \cite{jackson04} and the upcoming BRITE-Constellation of
microsatellites \cite{weiss07}.






\ack
The project "Stars: Central engines of the evolution of the Universe",
carried out at Aarhus University and Copenhagen University, is supported
by the Danish National Science Research Council.
HB is also supported by the Australian Research Council.


\bibliography{bruntt_helas2}

\end{document}